\documentclass[prl,floatfix,reprint]{revtex4}
\usepackage{graphicx,amssymb,amsmath}

\usepackage{subfigure}
\usepackage{subfig}
\begin{document}


\title{Multiplex networks in metropolitan areas: generic features and local effects}


\author{Emanuele Strano$^1$}
\author{Saray Shai$^{2,3}$}
\author{Simon Dobson$^2$}
\author{Marc Barthelemy$^{4,5,}$\footnote{Corresponding author (marc.barthelemy@cea.fr)}} 

\affiliation{$^1$ Laboratory of Geography Information Systems (LaSig), Polytechnic School of Lausanne (EPFL), Lausanne CH}
\affiliation{$^2$ School of Computer Science, University of St Andrews, St Andrews, Scotland UK}
\affiliation{$^3$ Department of Mathematics, University of North Carolina, Chapel Hill US}


\affiliation{$^4$ CEA, Institut de Physique Theorique, Gif-sur-Yvette FR}
\affiliation{$^5$ EHESS, Centre d'Analyse et de Math\'ematique Sociales, Paris FR}


\date{\today} 

\maketitle


Most large cities are spanned by more than one transportation system. These different modes of transport have usually been studied separately: it is however important to understand the impact on urban systems of the coupling between them and we report in this paper an empirical analysis of the coupling between the street network and the subway for the two large metropolitan areas of London and New York. We observe a similar behaviour for network quantities related to quickest paths suggesting the existence of generic mechanisms operating beyond the local peculiarities of the specific cities studied. An analysis of the betweenness centrality distribution shows that the introduction of underground networks operate as a decentralising force creating congestions in places located at the end of underground lines. Also, we find that increasing the speed of subways is not always beneficial and may lead to unwanted uneven spatial distributions of accessibility. In fact, for London -- but not for New York -- there is an optimal subway speed in terms of global congestion. These results show that it is crucial to consider the full, multimodal, multi-layer network aspects of transportation systems in order to understand the behaviour of cities and to avoid possible negative side-effects of urban planning decisions.


\section{Introduction}

In the last decade, urban transportation networks have been largely studied by means of spatial networks~\cite{Bar_rep} in order to understand aspects of urban systems and their evolution. These studies comprise the morphology of street networks~\cite{Jiang:2004,Roswall:2005,Marshall:2006,Porta:2006,Porta:2006b,Lammer:2006,Crucitti:2006,Cardillo:2006,Xie:2007,
Jiang:2007,Masucci:2009,Chan:2011,Barthelemy:2008,Courtat:2011,strano_EPB_2013,Louf:2014}, their evolution~\cite{Levinson:2011,Strano:2012,Barthelemy:2013,Porta:2014}, and their relationships with socio-economical indicators~\cite{porta,porta2}. In parallel, there are also studies on the structure of subway networks~\cite{Vito_boston,Derrible:2009,Derrible:2010}, their evolution~\cite{Roth}, and their robustness~\cite{Derrible:2010b,costa_EPL_2010}. However, these networks are \emph{not} independent, and the important result in~\cite{Buldyrev} showed that  coupling between networks can be critical and can affect the global behaviour of a system. 
It is in this context that multilayer (or multiplex) networks~\cite{Kivela,Shai,DeDomenico:2013,Bianconi} are studied and provide the convenient conceptual framework. A few recent studies considered the impact of the multilayer structure on various general processes~\cite{Gomez:2013,Domenico:2014}, and specifically in the case of transportation networks~\cite{Bar_prl,Gallotti:2014}.  Multilayer networks offer a good theoretical framework for understanding how interconnected transportation networks are shaping cities and how they may affect their operation. Moreover, given the increasing interest in urban systems, an empirical study of the impact of the multilayer structure of transport systems on mobility appears both crucial and timely.

In this study we consider the mutually connected underground and street networks in the large metropolitan areas of Greater London and New York City, and explore how their coupling affects their global properties. In particular, we analyse the effect of varying the subway speed and show that increasing it can lead to unexpected counter-effects. Our analysis focuses on three main network features and findings: i) the behavior of quickest paths at the city scale; ii) the local outreach and the urban horizon; and iii) the spatial distribution of betweenness centrality. It is important to stress that studies of urban transportation networks have important implications for urban policies and private investment, and in general play an important role in the urban planning chain. In fact, inter-modality transportation efficiency and simulations have been extensively studied in the transportation engineering literature~\cite{Farahani_2013}, where the typical supply-demand approach prevails but where the analysis of topological properties of networks is almost wholly neglected and where the different transportation modes are often treated separately. One goal in this study is to shift the focus onto this topological coupling aspect of transportation network design: we show this to be extremely relevant, and suggests that the multilayer network view of these systems should be integrated into elaborated models of urban planning. 

\section{Data and network construction}

\begin{figure*}[ht]
\begin{center}
\includegraphics[width=0.9\textwidth]{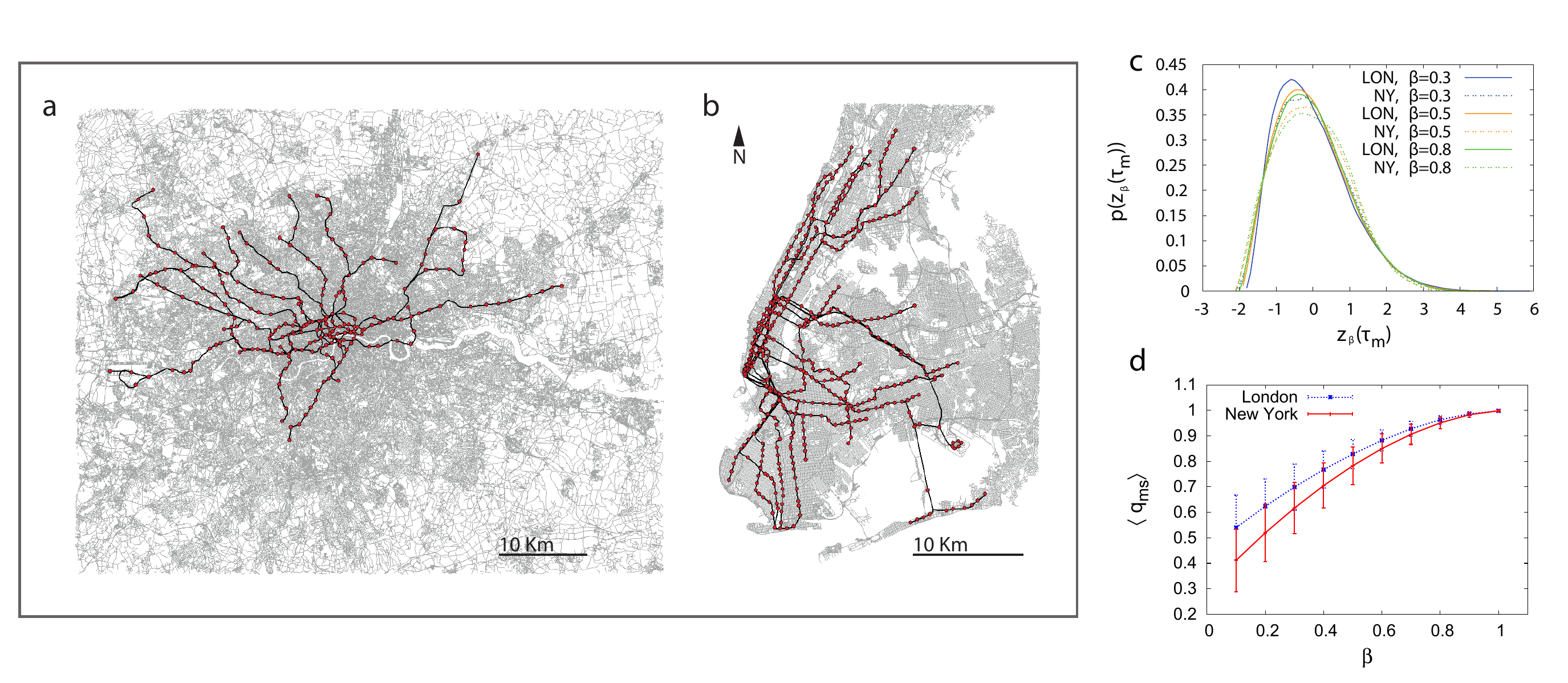}
\end{center}
\caption { (a,b) The spatial extent of the two metropolitan areas considered here. Note that the Greater London area is not covered by the underground system, in contrast with New York where most areas are connected by the subway.
(c) Distribution of normalised quickest path times  (computed in the multilayer system) $z_\beta(\tau_m) = \frac{\tau_m - \langle \tau_m \rangle}{\sqrt{Var(\tau_m)}}$. (d) The quantity $\langle q_{ms} \rangle$ averaged over all nodes as a function of $\beta$ (the error bars indicate here the dispersion around the average). The average ratio of travel times with and without the subway layer is typically of order $0.5$ and does not vary much with $\beta$. (The figures 1a,b were created using vectorial map extracted  in ArchMap environment and assembled with Adobe Illustrator.)}
\label{fig_1}
\end{figure*}

Using data from Open Street Map ~\cite{OSM}, we construct both the street and the subway networks for London (UK) and New York City (USA). We downloaded data on street networks and underground in geo-referenced vectorial format from 
Open Street Map, which contains detailed streets and rail tracks networks, including train depots and double tracks. (The rationale behind the geographical extent of these networks is to include the full underground systems and surrounding street networks.) 
In addition, a series of automatic and manual topological cleaning operations were needed in order to extract consistent and usable graphs.The size and geography of the two cities is clearly different as we can observe it in Fig.~\ref{fig_1}a,b. 

We thus obtained the weighted graph $G_{\text{s}} = (V_{\text{s}}, E_{\text{s}})$ of the connected street network in its ``primal'' representation, with nodes being street junctions and edges representing the street segments connecting them, and the weights given by the street length $W=street length (m)$. Similarly, we obtained the connected underground network $G_{\text{u}} = (V_{\text{u}}, V_{\text{u}})$ with nodes representing underground stations and links connecting successive stations on the same line, and weighted by the length of the line segment $W=link length (m)$. From a theoretical point of view, the interdependent or multilayer~\cite{Kivela} network, $G_{\text{multi}}$ is defined as the union of these two networks. Here we have subway stations and road intersections that we consider to be different nodes. Underground stations are accessible from more that one access on the street, but for the sake of simplicity we construct the multilayer network by connecting each underground station to its closest street junction only (a simplification that won't change the structure of quickest paths). 
In order to create the adjacency lists we used a combination of Python scripts, ArchMap geo-tools (ArchGIS 10.2) and \textit{ad hoc} manual corrections. Tools have been set to remove links redundancies, to correct the topology of the networks, and to create the proximity matrix between street nodes (street junctions) and underground nodes (stations). The scripts have been corroborated with a full check of the data and corrections of the topology in editing sessions under ArchMap environment (the computation of the various statistical measures have been done under Python environment using NetworkX library and the maps have been produced using ArchGIS 10.2).

\section{The generic nature of quickest paths}

New York is composed of two large and almost disconnected components with the underground systems covering a similar spatial extent and carving-up the different boroughs. London instead presents -- at a large scale -- a typical radiocentric urban structure with the underground systems connecting satellite districts and peripheries to the urban core.  Differences both in size and geography between these cities are also reflected by basic network descriptors shown in Table~\ref{table:table1}. 
\begin{table*}[ht]
    \begin{tabular}{ | l | l | l | l | l | l | l | l | l | l | l |  }
    \hline
    Case      & N & M & Total length (km) &
  $\overline{l}_{ij}^{geo}$ (km) &   $\overline{l}_{ij}^{topo}$ &  $Diam^{geo}$ (km) & $Diam^{topo}$ & $ \sqrt{A}$(km) & Node density (km$^{-2}$)\\ 
\hline
    London streets  & 324536 & 427920 & 34493.73 & 25.83 & 178.16 & 89.31 &368 & 73.68  & 59.78\\ 
    London tube & 263 & 296 & 385.98 & 18.55 & 14.26 & 60.3 & 42  & -  &0.048  \\ 
    London multi & 324799 & 428479 & 34886.52 & 25.78 & 96.16 & 89.27 & 288  &- &-  \\\hline
    New York street & 68417 & 112827 & 12153.81 & 17. 94 & 106.64 & 55.18 & 278 & 34.93  &56.07  \\
    New York subway & 454 & 489 & 416.12 & 18.87 & 19.10 & 57.28 & 62  & -  &0.37  \\
    New York multi & 68871 & 113770 & 12579.44 & 17.91  & 54.45 & 55.18 & 205&- &-   \\\hline
    \end{tabular}
\caption{The number of nodes $N$, the number of links $M$, the cost defined as the total length of all links, the average geographical and topological shortest path length, $\overline{l}_{ij}^{geo}$ (km) and $\overline{l}_{ij}^{topo}$ respectively; the maximum geographical and topological shortest path length, $D^{geo}$ (km) and $D^{topo}$ respectively; the density of streets $\rho_s = \frac{N_s}{A}$ and the density of underground stations $\rho_u = \frac{N_u}{A}$, where $A$ is the area of each city.
}
\label{table:table1}
\end{table*}

For both cities, the (spatial) diameter of the multiplex is essentially dominated by the street network. We also observe that the topological diameter of the multiplex is lower than the street layer, thanks to the subway structure allowing for topologically shorter paths. The efficiency of the subway is however also due to its speed which is in general larger than that of overground modes such as private cars, taxis, or buses. In order to reflect this, we introduce a parameter $0 < \beta \leq 1$ that describes the ratio of speeds in both systems, similarly to the theoretical analysis proposed in~\cite{Bar_prl}. This parameter $\beta$ measures the travel cost in time units associated to the underground links. This means that the number of time units taken to traverse an underground link of length $l$ meters is $\beta l$, which is $\frac{1}{\beta}$ times faster than the time taken to traverse the same length on the street network. Thus a smaller $\beta$ corresponds to a faster underground speed, as compared with the speed on the street network. The introduction of this parameter allows us to study the properties of the multilayer system as a function of underground speed. Na\"ively one could expect that the system as a whole will be more efficient for faster subways, but we will show here that it is not always the case and that in some cases we can observe an optimal value for $\beta$. Finally $\beta$ can be measured empirically, and we obtain for London $\beta_{London}\approx 0.48$ and a slightly larger value for NY $\beta_{NYC}\approx 0.55$.

We denote by $\tau_s(i,j)$ the travel cost (\textit{i.e.}, the number of time units) of the quickest path between street nodes $i,j \in V_\text{s}$, and by $\tau_m(i,j)$ the cost of the quickest path between $i$ and $j$ in the multilayer network (\textit{i.e.}, a path which can traverse \emph{both} street \emph{and} underground links). The normalised quantity

\begin{equation}
  z_\beta(\tau_m) = \frac{\tau_m - \langle \tau_m \rangle}{\sqrt{Var(\tau_m)}}
\end{equation}

\noindent displays a behaviour that is roughly constant for $\beta$ larger than $0.2-0.3$, as shown in Fig.~\ref{fig_1}c, demonstrating that the effect of $\beta$, is essentially contained in the average and variance of $\tau_m$. This is a rather surprising result, given that the two cities display many geographical and structural differences. The cost $\tau_m(i,j)$ between nodes i and j can be written as

\begin{equation}
  \tau_m(i,j)=\sum_{e\in P(i,j)}\tau(e)
\end{equation}

\noindent where the sum is over all links $e$ that belong to the quickest path $P(i,j)$ and where $\tau(e)$ is the cost on this link. (We neglect inter-modal change costs in this simple argument.) If the path is long enough, and if the random variables $\tau(e)$ do not display long-range correlations and are not broadly distributed, the central limit theorem  applies and the distribution of the
$\tau_m$ follows a Gaussian distribution in a certain range.  There are obviously deviations observed for small values of $\beta$ coming from the fact that the paths' durations become very heterogeneous depending on the proximity of their origin or destination to subway stations. In this respect, a very high relative subway velocity enhances spatial differences in the city and may lead to an uneven distribution of accessibility, a fact that will be confirmed below with the local outreach analysis.

We also compute the average ratio between the travel costs from $i$ to other street nodes through the multilayer network and through the street network, defined as

\begin{equation}
  q_{ms}(i) = \frac{1}{N_{\text{s}} - 1} \sum_{j \in V_{\text{s}}} \frac{\tau_m(i,j)}{\tau_s(i,j)}
  \label{eq:qms}
\end{equation} 

\noindent where $N_{\text{s}}$ is the number of street nodes. The larger this ratio, the larger the effect of the underground on travel costs. We see in Fig.~\ref{fig_1}d that typical values are of order $0.5$ for both cities and that the effect of $\beta$ is rather weak: a decrease from $\beta=1$ to $\beta=0.5$ leads to a decrease of $\langle q_{ms}\rangle$ of order $20\%$.  In addition, it seems that the effect of subways in London is less important than in New York, which is probably due to the lesser extent of the subway in the Greater London area.

A central quantity for describing the importance of inter-modality is given by

\begin{equation}
  \lambda(i,j) = \frac{\sigma_{i,j}^{\text{multi}}}{\sigma_{i,j}} 
\end{equation}

\noindent where $\sigma_{i,j}$ is the total number of shortest paths between $i$ and $j$ (using either one or two networks), and $\sigma_{i,j}^{\text{multi}}$ the number of paths using edges of both networks at least once. It characterizes the importance of multi-modality for the path from $i$ to $j$.  If we sum over all possible destination nodes $j$, we can quantify the added value of the interlayer coupling to the reachability of nodes, and obtain the interdependency~\cite{Bar_prl} of a street node $i \in V_{\text{s}}$ defined as

\begin{equation}
  \lambda(i) = \frac{1}{N_{\text{s}}} \sum_{j \in V_{\text{s}}} \frac{\sigma_{i,j}^{\text{multi}}}{\sigma_{i,j}} 
  \label{eq:lambda}
\end{equation}

\begin{figure}[h]
\subfigure{\includegraphics[width=.4\textwidth]{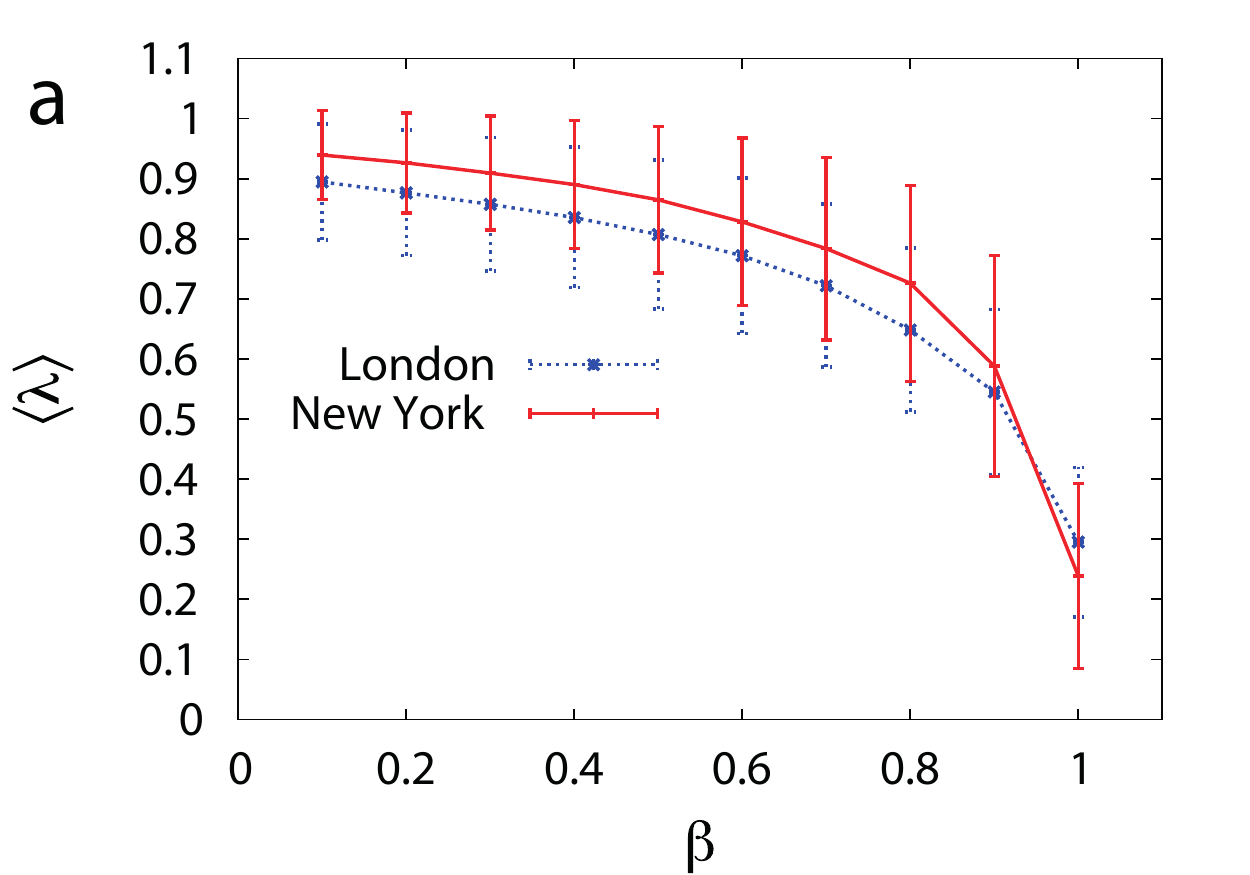}}
\subfigure{\includegraphics[width=.4\textwidth]{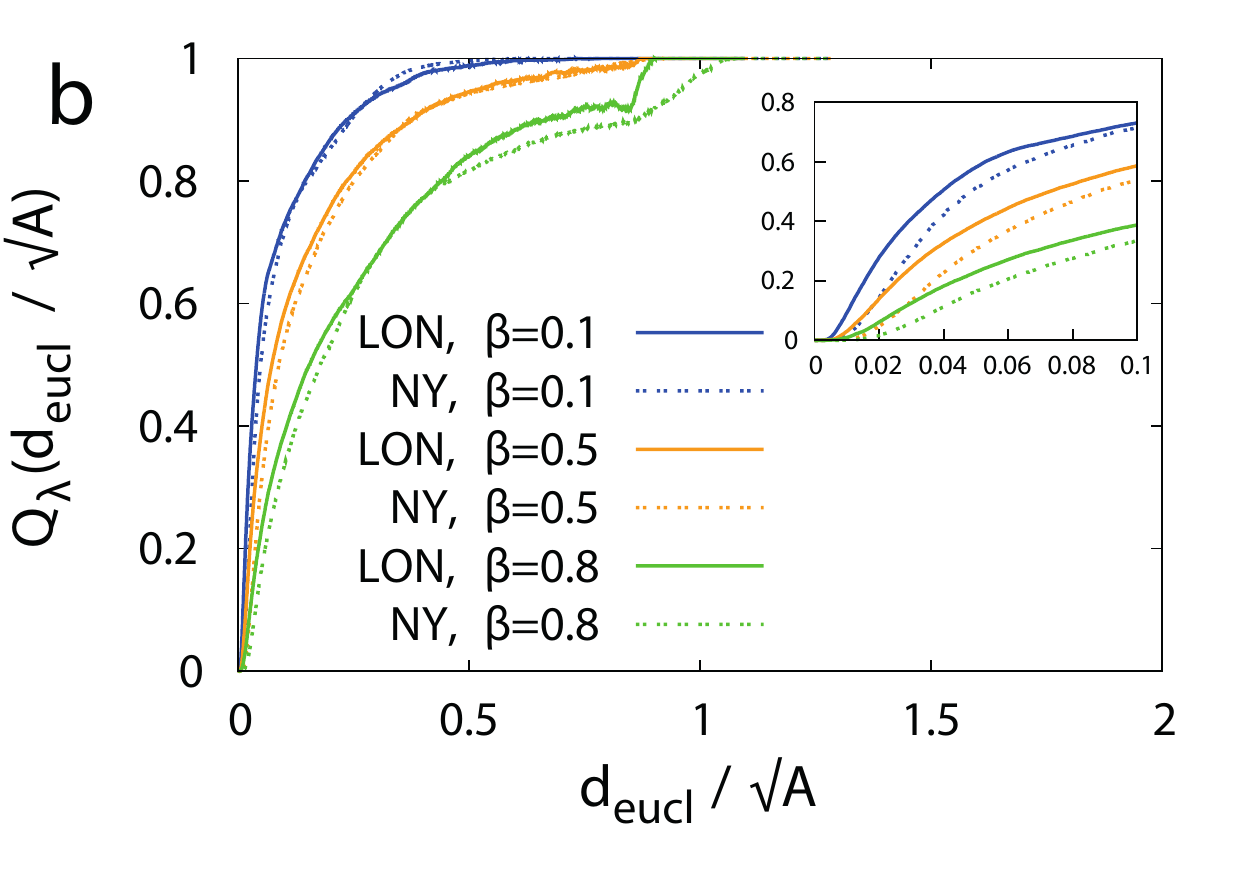}}
\caption{(a) Average interdependence, $\langle \lambda \rangle$ as a function of $\beta$. (b) Normalised interdependence profile computed for different values of $\beta$. Both cities exhibit a similar behaviour despite very different geographical structures.} 
\label{fig_3}
\end{figure}

\noindent (Note that a similar measure has been used in the transportation design literature under the name of \emph{inter-modal connectivity}~\cite{Farahani_2013}.) In order to understand the effect of scale on the interdependence, we also define the interdependence profile as

\begin{equation}
  Q_{\lambda}(d) = \frac{1}{N(d)} \sum_{\substack{i, j \in V_{\text{s}} \\ d_e(i,j) = d}} \lambda(i,j)
  \label{eq:profile}
\end{equation}

\noindent where $d_e(i,j)$ is the Euclidean distance between $i$ and $j$ and $N(d)$ is the number of pairs of nodes at Euclidean distance $d$. In Fig.~\ref{fig_3}a we show the average interdependence among all street nodes as a function of $\beta$ and the resulting interdependence profile Fig.~\ref{fig_3}b. 

We see from these figures that, in both cities, the existence of the underground has a very large impact. 
For example, for $\beta = 0.8$ we obtain $\lambda$ around $0.7$, meaning that even when the underground is only $1.25$ times faster than the street network, already about $70\%$ percent of the quickest paths are going through the underground. A slight decrease of $\beta$ for $\beta$ close to one thus has a large impact on structure of quickest paths, while for smaller values of $\beta$, improving the subway speed does not bring a significant improvement of the structure of quickest paths. In both cities there is a sharp increase in $\lambda$ for small Euclidian distance, meaning that already for relatively short trips, it is worth ``hopping on'' to the underground. (Note that we neglect here waiting, walking, and connecting times which can significant~\cite{Gallotti:2014}.) The slope of the interdependence profile at small $d_{\text{eucl}}\simeq 0$ is increasing as $\beta$ is decreasing, suggesting that a slight increase in the underground speed could make the networks highly interdependent even at very small scales.

Both cities therefore display a remarkably similar behaviour over all these interdependency-related quantities (in particular, see Fig.~\ref{fig_3}(b)), suggesting here again a possible common behaviour for multiplex transportation network in cities. 
While further studies are needed in order to substantiate a claim of ``universality'', our results point to the possible existence of some kind of a statistical law of large numbers that applies to quickest paths in multiplex urban transportation networks. 

We note that it is not trivial that the central limit theorem applies here and it doesn't mean that the network topology is irrelevant. The fact that we can sum a large number of quantities, which are uncorrelated (a necessary condition for the central limit theorem to apply) comes from the specific structure of these transportation systems (spatial constraints for example certainly play an important role). In addition, more complex quantities (such as the interdependence for example) also display a large level of similarity for the two cities, a fact that cannot at this stage be simply related to a central limit theorem. These different results point to the potentially useful fact that actually few parameters seem to govern the behavior of these quantities, which could lead to many useful simplifications in more elaborated models that contain a large number of parameters.

\section{Local outreach and the urban spatial horizon}

\begin{figure}[h]
\subfigure{\includegraphics[width=.4\textwidth]{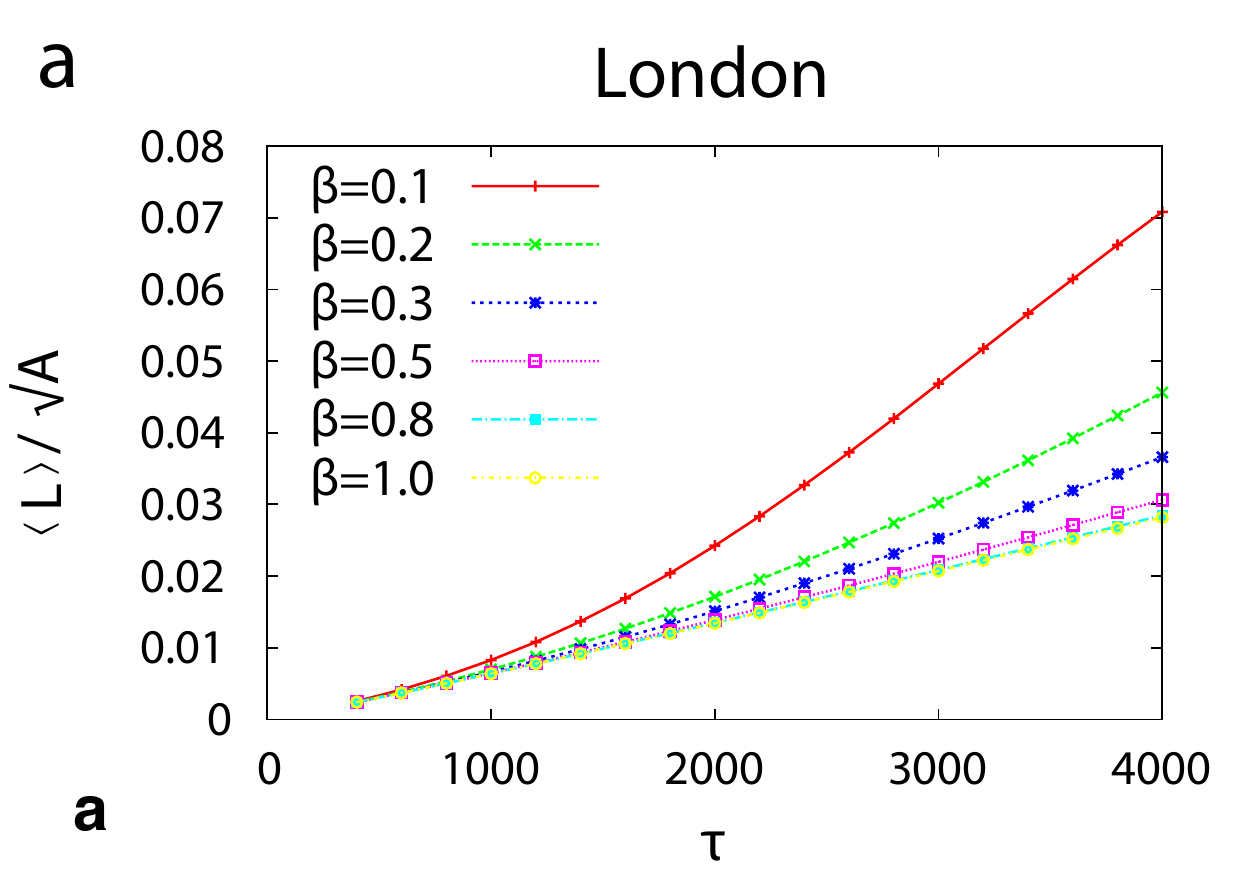}}
\subfigure{\includegraphics[width=.4\textwidth]{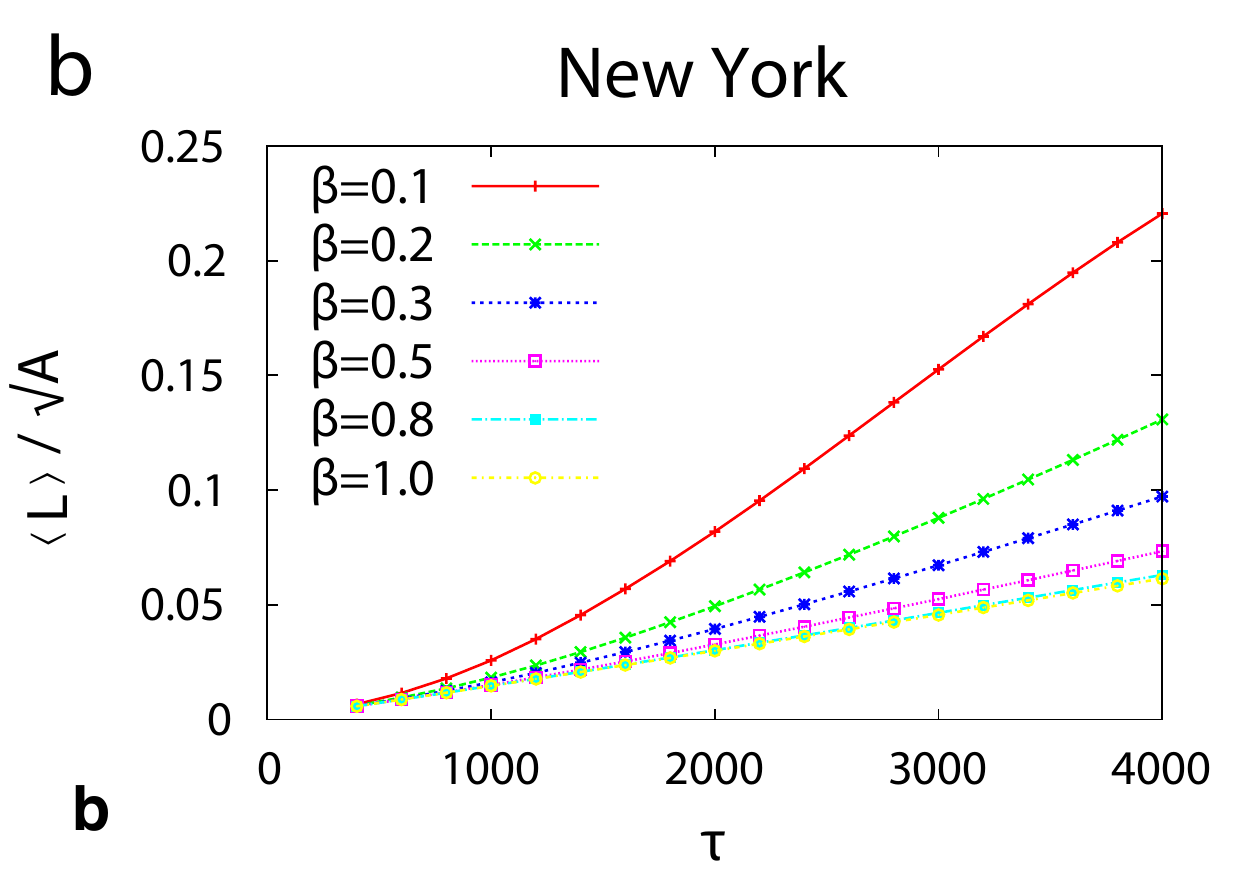}}
\caption{Average local outreach $\langle L \rangle$ normalised by the square root of area, for London (a) and New York (b).} 
\label{fig_4}
\end{figure}

The presence of a transportation mode such as a subway affects the overall performance of a city in terms of efficiency of transport and the accessibility of certain locations, but also has an important impact on how pairs of locations are connected. In order to measure this effect, we define the {\it spatial outreach} of a street node $i \in V_\text{street}$ as the average Euclidean distance from $i$ to all other street nodes that are reachable within a given travel cost, $\tau$:

\begin{equation}
L_\tau(i) = \frac{1}{N(\tau)} \sum_{j | \tau_m(i,j) < \tau}  d_e(i,j)
\label{eq:outreach}
\end{equation}

\noindent where $d_e(i,j)$ is the Euclidean distance between node $i$ and $j$, and $N(\tau)$ is the number of nodes reachable on the multilayer network within a given travel cost $\tau$. In Fig.~\ref{fig_4} we show the average local outreach as a function of the travel cost threshold $\tau$, which displays a non-linear behaviour due to the different speeds achievable in the two transportation modes. This provides support for a general effect already known: for longer trips, faster transportation modes are used (see for example~\cite{Gallotti:2014} for the UK case).

For New York, the unit of time is given by the average car speed on the street network which is of order $15.6$km/h (see for example~\cite{NYCforum}). Rescaling $\tau$ by this velocity, we then obtain an effective maximum speed (for $\beta=0.1$) of order $30$km/h (close to the $28$km/h discussed in~\cite{NYtimes}). For London, the same calculation with an average car speed of $16$km/h (see~\cite{tfl}) yields an effective maximum speed of order $21$km/h. (This difference in speeds is due to the areas considered, as New York is almost entirely covered by the underground network.)

\begin{figure*}[ht]
\includegraphics[scale=0.9]{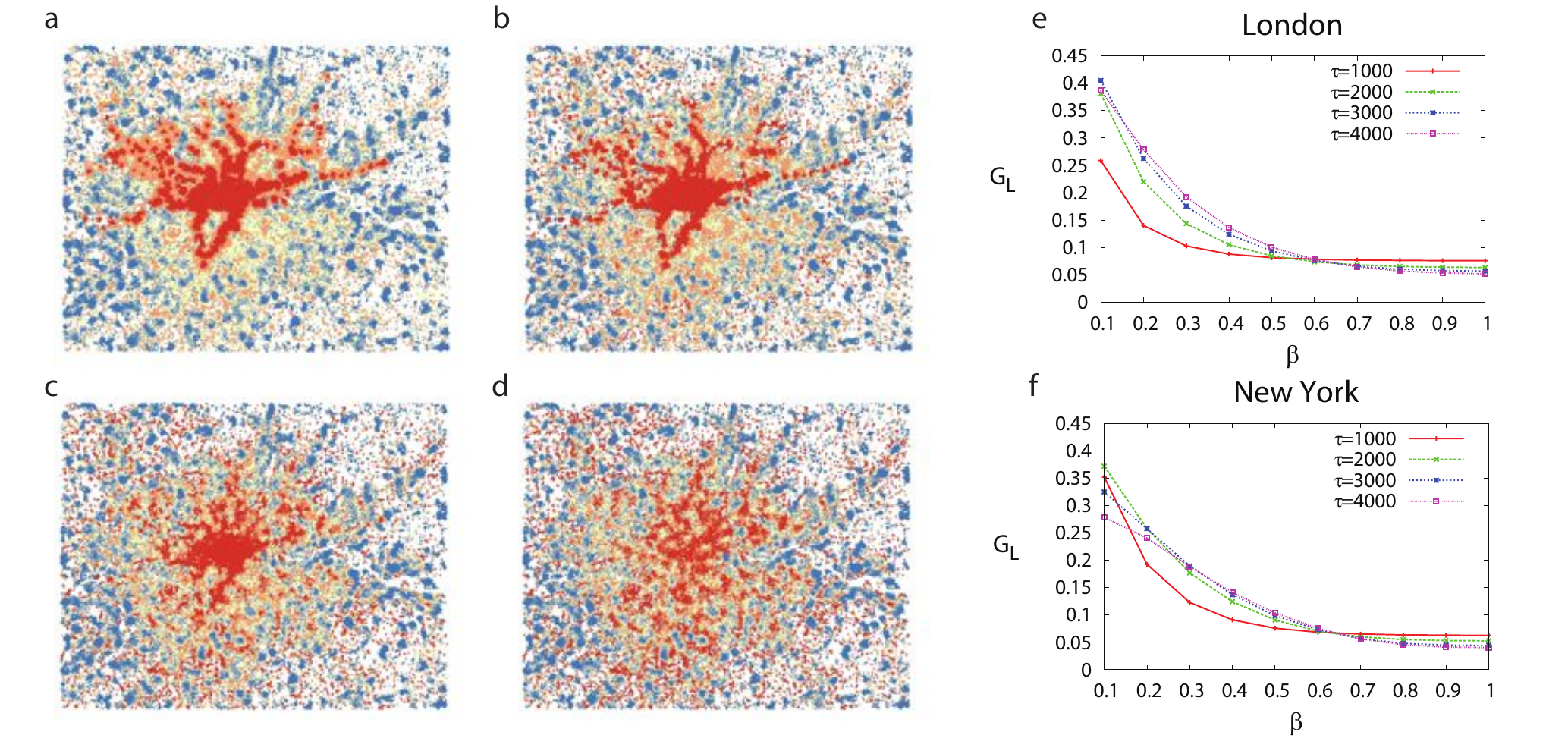}
\caption{Maps showing for London the spatial distribution of the local outreach defined in Eq.~\ref{eq:outreach} for a travel cost $\tau= 2000$ and speed ratio ( a) $\beta$=0.1, (b) $\beta$=0.7, (c) $\beta$=0.4 and (d) $\beta$=1. As the underground's speed is increasing compared to the speed of the street network (\textit{i.e.}, decreasing $\beta$) the nodes having a high local outreach are concentrated at the center along the underground network, where the underground is most accessible. Theses figures were created using ESRI ArchMap 10.1 and Adobe Illustrator. (e, f) The graphs show the Gini coefficient of the local outreach versus $\beta$ for both cities.} 
\label{fig_5}
\end{figure*}

As shown is Fig.~\ref{fig_5},a,b,d,e as $\beta$ decreases, the nodes having a high local outreach are concentrated close to underground stations where the underground is the most accessible, and the graph consisting of high-outreach nodes (red nodes on the map) becomes less fragmented. 

In other words, as the underground becomes faster, a continuous area of high-outreach nodes emerges (the \emph{commutable zone}) in the city centre and around the nodes of the underground network, implying that a person can travel from this area to faraway places (large Euclidean distance) at a small travel cost $\tau$. The location of this highly accessible zone cluster from a dispersed configuration, as in Fig.~\ref{fig_5}d, to a centralised one, as in Fig.~\ref{fig_5}a, which shows a centralisation effect due to the accessibility provided by the underground. The dispersion of the local outreach also displays a very interesting result demonstrated by its Gini coefficient $G_L \in [0,1]$. Indeed, in Fig.~\ref{fig_5}e,f we see that for both cities for $\beta>0.5$ the accessibility is distributed almost uniformly amongst all the places in the cities, while for smaller $\beta$ (faster underground) the shift to an uneven distribution of accessibility is clear. This result suggests that transportation policies that focus on increasing the speed on a single travel modality may lead to the undesirable spatial heterogeneity in the accessibility of different locations.

\begin{figure*}[ht]
\subfigure{\includegraphics[width=.35\textwidth]{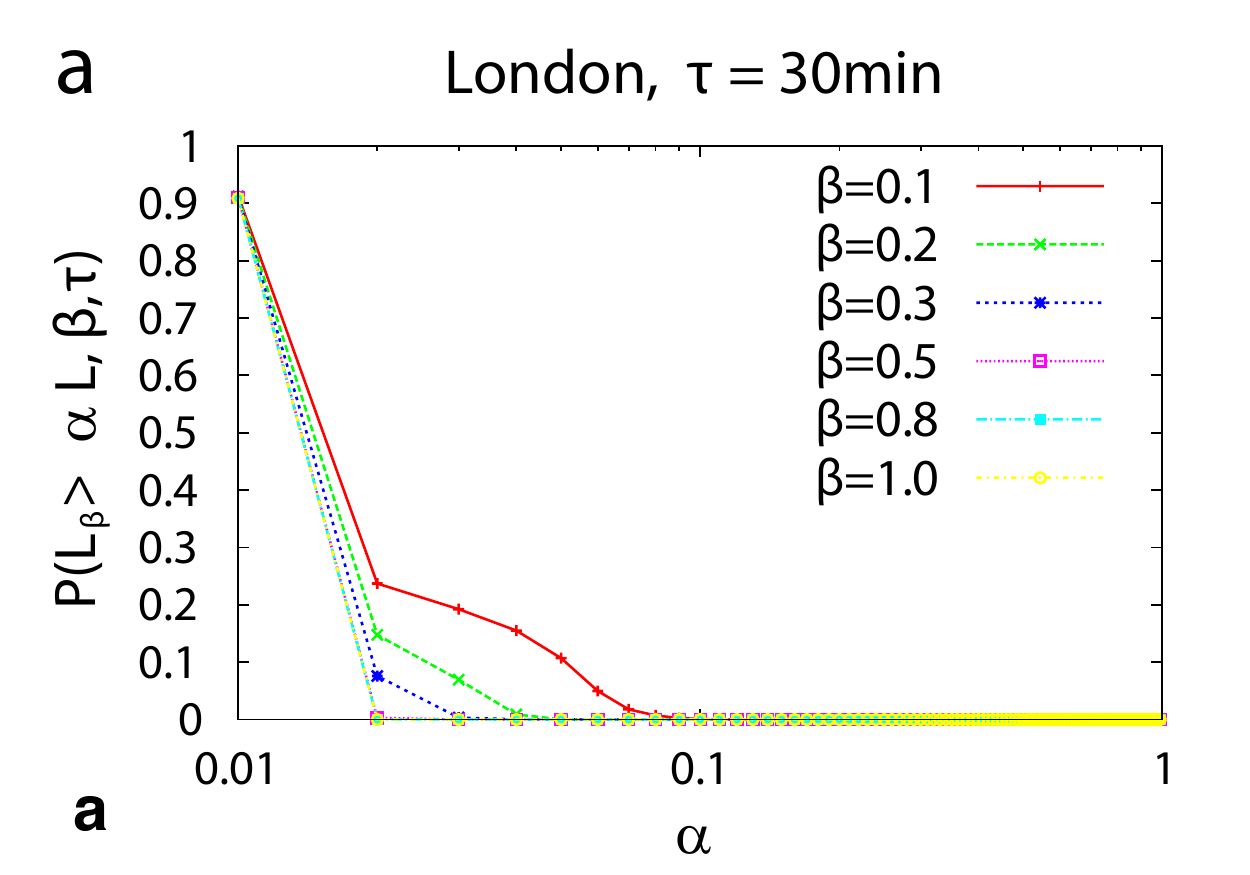}}
\subfigure{\includegraphics[width=.35\textwidth]{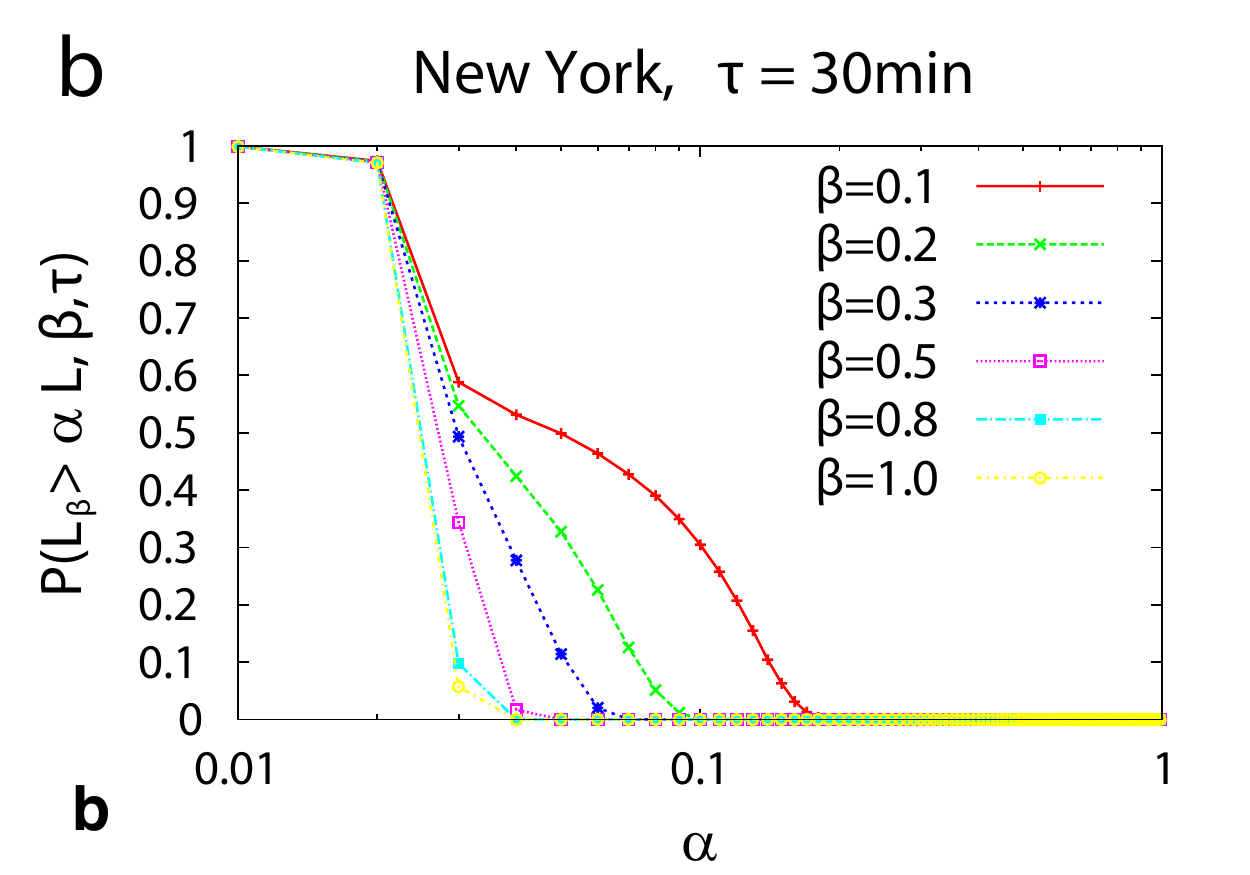}}
\subfigure{\includegraphics[width=.35\textwidth]{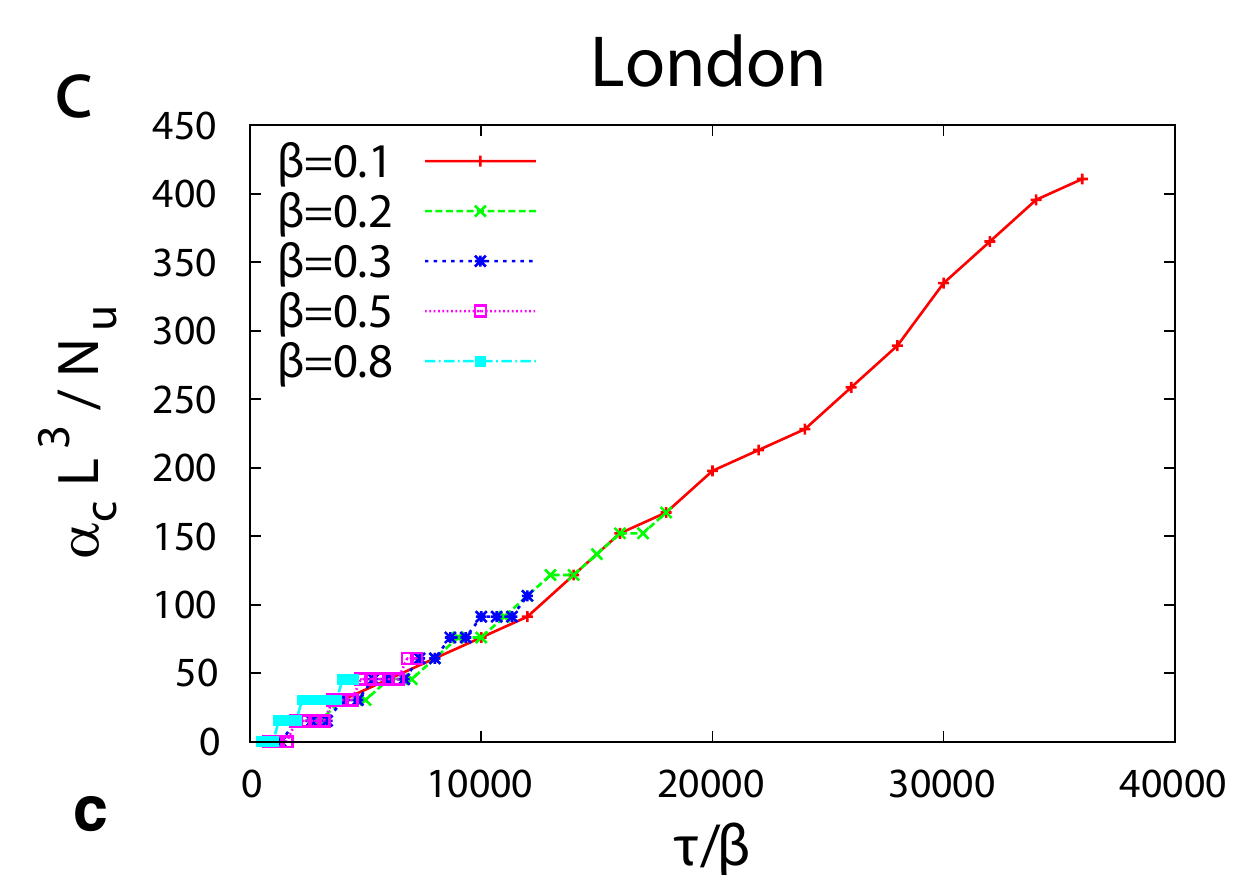}}
\subfigure{\includegraphics[width=.35\textwidth]{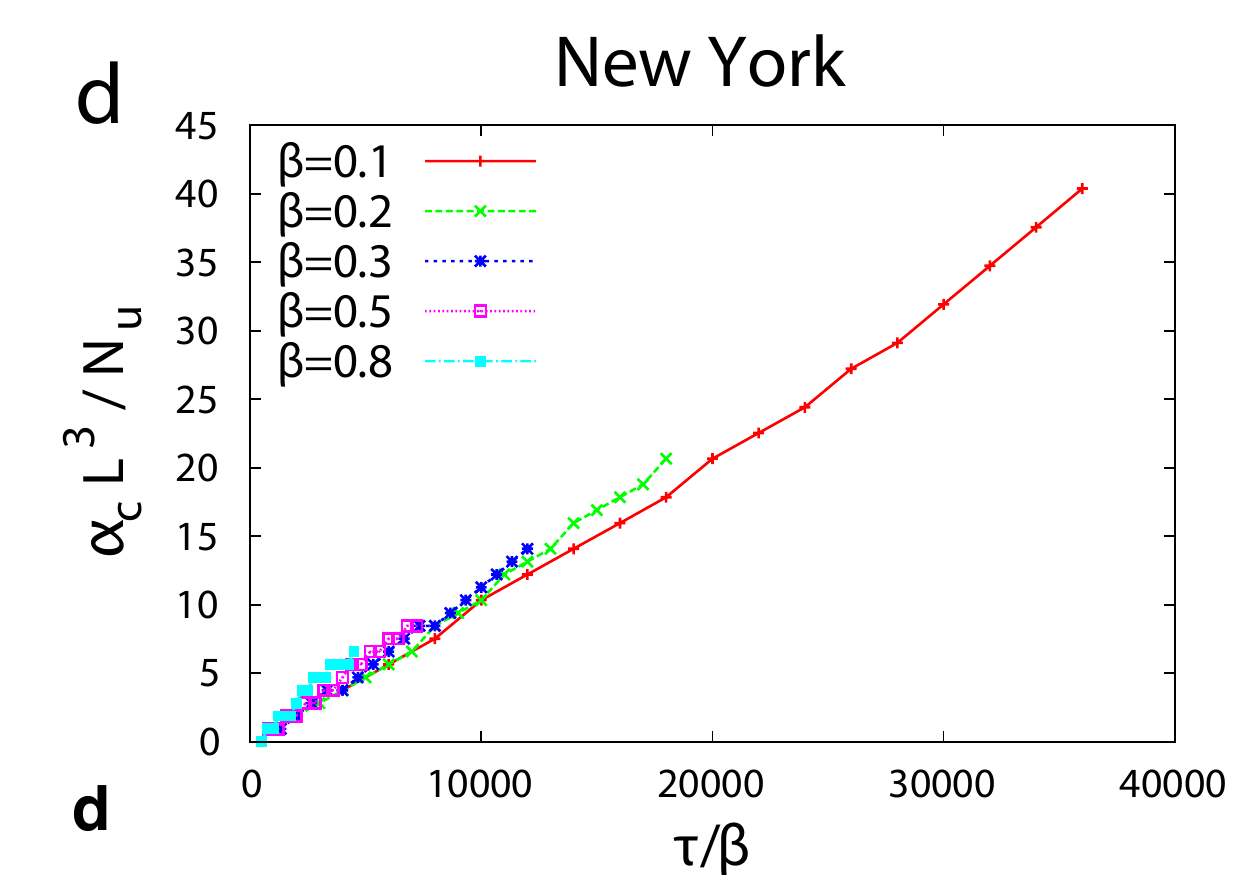}}
\caption{(a)-(b) Fraction of nodes with a local outreach larger than $\alpha \sqrt{A}$ for London (a) and New York (b). (c)-(d) when both axis are rescaled according to Eq. 8, we observed a data collapse with all curves collapsing on a single one.} 
\label{fig:pL}
\end{figure*}

We show in Fig.~\ref{fig:pL} the probability that the outreach is larger than a certain fraction $\alpha L$ of the size of the city, and we observe the existence of a threshold $\alpha_c$. 

The existence of a threshold less than one means that, for given values of $\beta$ and $\tau$, there is a maximal value $L_m$ for the outreach. We can estimate the value of $L_m$ by using a simple argument: the maximum value is reached when the path is ``essentially'' made on the quickest transportation mode, the subway. This transportation mode has a velocity given by $v/\beta$, and the probability that a station is within reach (in circle of radius $d_0$ corresponding to the typical walking distance to reach the subway) is

\begin{equation}
  p=\rho_u \pi d_0^2
\end{equation}

\noindent where $\rho_u=N_u/A$ is the density of subway station ($A=L^2$ is the area of the city and $N_u$ is the number of subway stations). The maximal outreach $L_m$ is then given by

\begin{equation}
  L_m=\frac{N_u}{L^2}\pi d_0^2\frac{v}{\beta}\tau
\end{equation}

\noindent and $\alpha_c=L_m/L$ is then given by

\begin{equation}
  \alpha_c=\frac{N_u}{L^3}\pi d_0^2 v\frac{\tau}{\beta}
\end{equation}

\noindent This last equation shows in particular that the quantity $\alpha_cL^3/N_u$ should increase linearly with $\tau/\beta$, with a constant of proportionality depending on geometry the city, and we observe that this scaling in is agreement with simulations (see figures~\ref{fig:pL}c, d). In particular, we see that the slopes for London and New York are different: the ratio of the constant pre-factors is about $10$, suggesting that the subway system in London is more efficient in terms of the outreach that can used as a measure of the ``urban horizon''.

\section{The geography and distribution of urban centrality}

Betweenness centrality (BC)~\cite{Freeman} is one of the important quantities in complex networks, and in street networks in particular~\cite{Crucitti:2006}. It quantifies the importance of a node as being the amount of traffic going through it, assuming uniform demand where the traffic between all pairs of nodes is the same. This quantity is very relevant in urban systems: in particular, it is correlated with the locations of shops and other micro-economic activity~\cite{porta,porta2}, urban growth~\cite{Strano:2012,Barthelemy:2013}, and land-use intensity~\cite{WanWAP11}. 

In the case of car traffic and congestion, the absence of detailed traffic models or mobility data leads us to use the BC in order to identify the \emph{potentially} congested locations and the effects of spatial structure on the shortest path structure.
Even if we know that the assumptions used in the BC calculation can lead to some inacurracies \cite{Gonzalez_bet_2013}, it is the simplest proxy that contains some level of information about real traffic. We thus explore in this section the spatial distribution of BC in the street network and how it is affected by the underground system. The BC of a street node $v \in V_{\text{s}}$ in the street network is defined as 

\begin{equation}
  bc_\text{s}(v) = \frac{1}{(N_{\text{s}} -1)(N_{\text{s}} - 2)} \sum_{i, j \in V_{\text{s}}} \frac{\sigma_{i,j}^{\text{street}}(v)}{\sigma_{i,j}^{\text{street}}}
  \label{eq:bcstreet}
\end{equation}

\noindent where $\sigma_{i,j}^{\text{street}}$ is the number of quickest paths between $i$ and $j$ in the street network, of which $\sigma_{i,j}^{\text{street}}(v)$ goes through street node $v$. Similarly, we define the betweenness centrality of a street node $v \in V_{\text{s}}$ in the multilayer network as

\begin{equation}
  bc_\text{m}(v) = \frac{1}{(N_{\text{s}} -1)(N_{\text{s}} - 2)} \sum_{i, j \in V_{\text{s}}} \frac{\sigma_{i,j}^{\text{multi}}(v)}{\sigma_{i,j}^{\text{multi}}}
  \label{eq:bccoupled}
\end{equation}

\noindent where $\sigma_{i,j}^{\text{multi}}$ is the number of quickest paths between $i$ and $j$ in the multilayer network, of them $\sigma_{i,j}^{\text{multi}}(v)$ goes through street node $v$.

\begin{figure*}[ht]
\includegraphics[width=.9\textwidth]{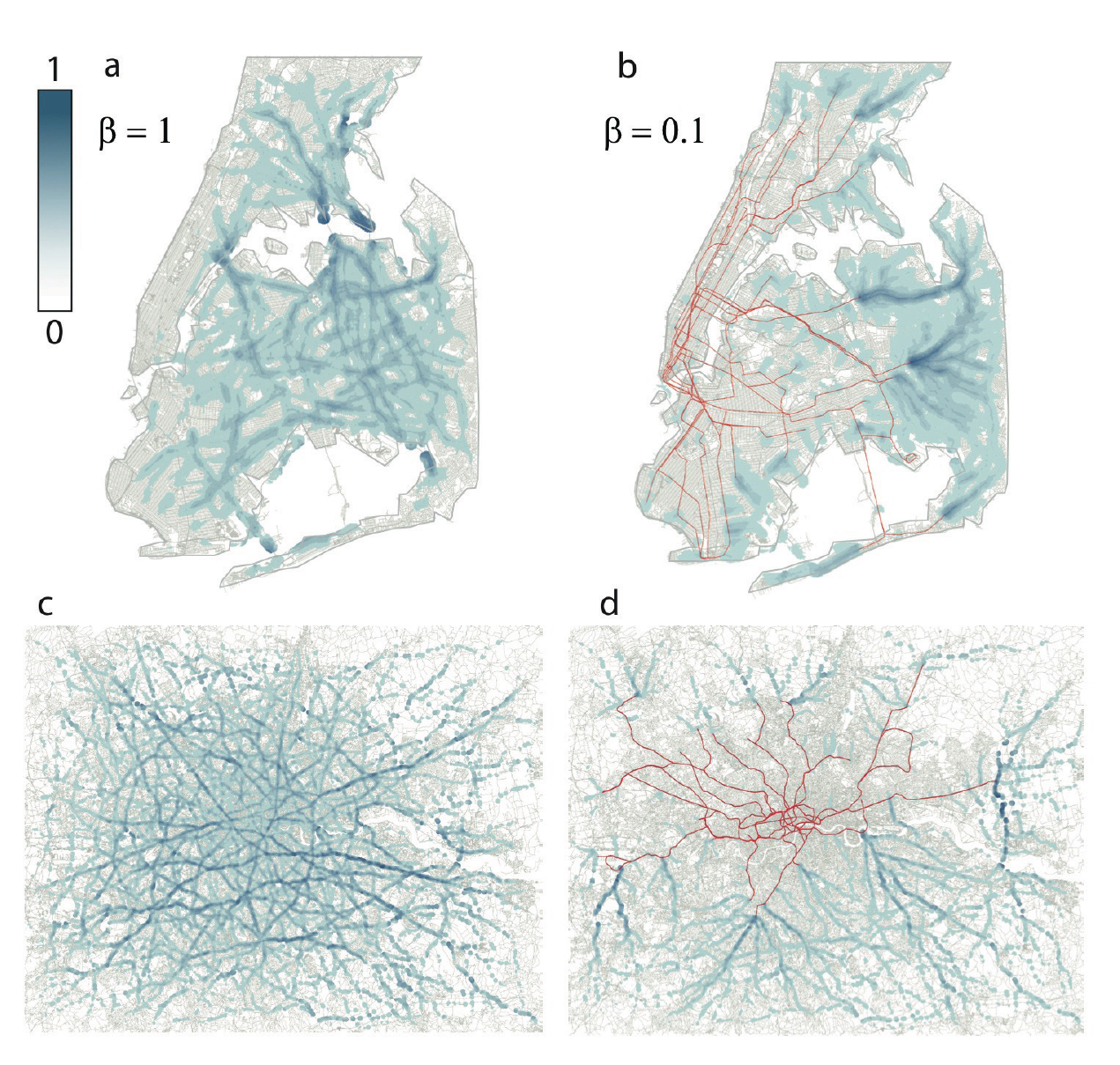}
\caption{The spatial distribution of BC on the New York (top) and London (bottom) street network for different values of $\beta$. We observe a clear crossover from congested road locations for $\beta=1$ to ``focal points'' of the underground system for small $\beta$ (The maps have been obtained using the classical interpolation method Inverse Distance Weigh (IDW) on the street junctions with Z = BC). This figure was created using ESRI ArchMap 10.1 and Adobe Illustrator.}
\label{fig:mapsbc}
\end{figure*}

We can then observe how the parameter $\beta$ impacts the mobility distribution and the geography of potentially congested areas. The maps in Fig.~\ref{fig:mapsbc}(a,b,c,d) show the BC spatial distribution for both cities computed on streets for $\beta =1$ (a,b) and $\beta =0.1$ (b,c). 

\begin{figure}[ht]
\includegraphics[width=.43\textwidth]{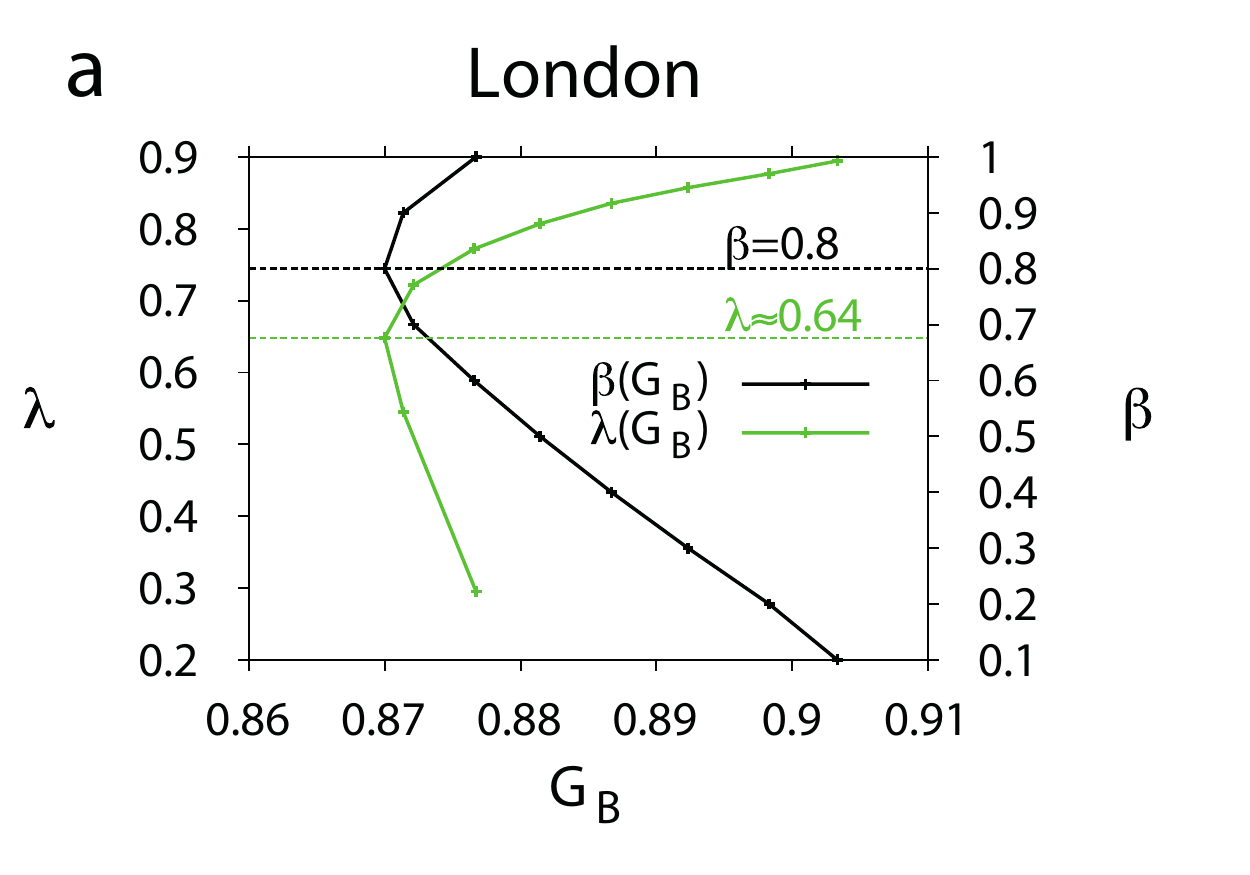}
\includegraphics[width=.4\textwidth]{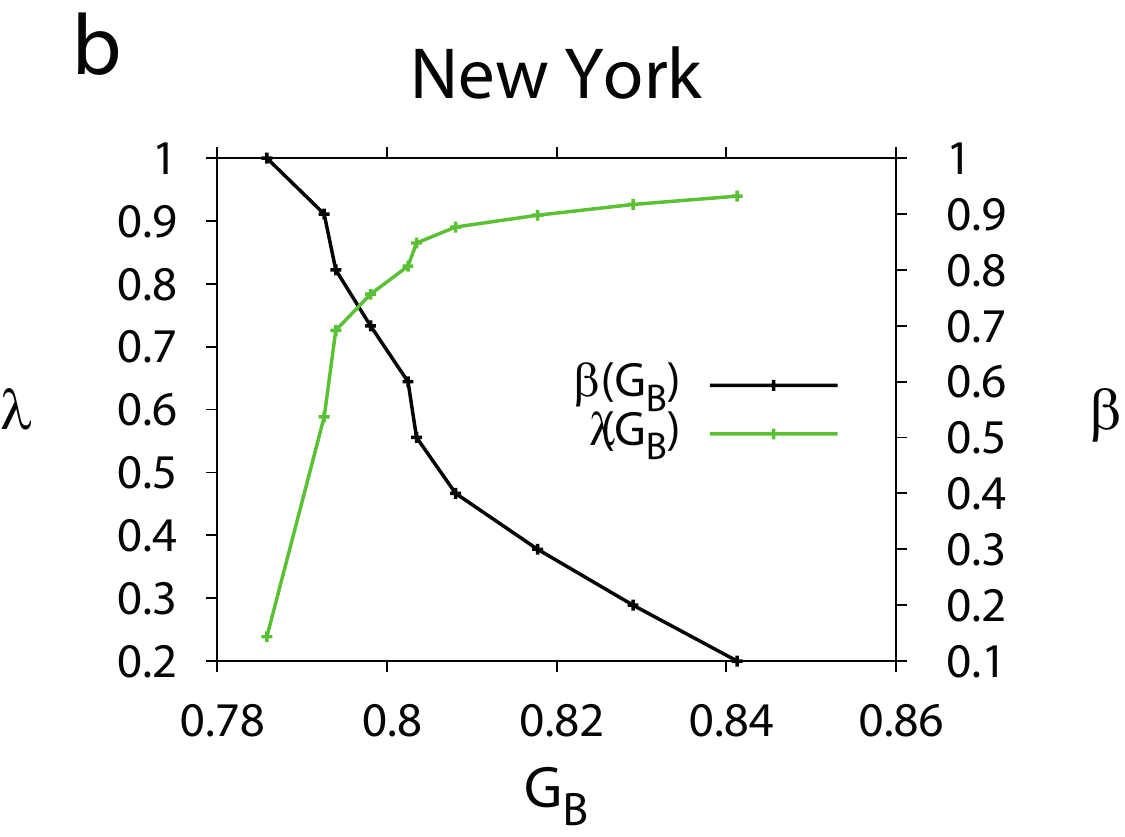}
\caption{We represent here for London (a) and New York (b) $\beta$ and interdependency $\lambda$ as a function of the dispersion of the multilayer BC (measured by the Gini coefficient $G_B$ for the BC).} 
\label{fig:gini}
\end{figure}

These maps clearly display a dramatic change in the spatial distribution of central places when introducing an underground system, shifting congestion from internal street routes and bridges to inter-modal places located at the terminal points of the underground networks, which presumably are used as entry/exit gates for suburban flows to reach core urban areas. Remarkably enough, in both cities, these places are located in urban areas that do not overlap with the underground system, thus possibly creating congestion in unexpected places. In other words, the introduction of underground networks operate as a decentralising force creating congestion in places located at the ends of underground lines and not, for example, in the city centre as one might expect referring to classical results on rewiring processes for chain or lattice networks~\cite{Bar_rep} in which BC is correlated to the distance to the gravitational centre. The statistical dispersion of BC can be measured by its Gini coefficient and also suggests that congested places always become more critical in the system as $\beta$ decreases. In fact, as shown in figure ~\ref{fig:gini}, the Gini coefficient of BC increases as the underground becomes more efficient (faster, decreasing $\beta$), meaning that a larger fraction of quickest paths use it; and the BC distribution is less homogeneous, making the system more fragmented and less resilient.  

Examining BC Gini as a function of $\beta$ and the interdependency $\lambda$ in London (Fig.~\ref{fig:gini}), we observe a non-trivial optimal value for $\beta$ for which flows are the most homogeneously distributed across street junctions. This minimum value for the Gini also corresponds to a minimum of the total congestion (see the appendix Total congestion).
In New York, however, there seem to be room for small $\beta$ and small congestion and the absence of a non-trivial optimum for New York suggests (as discussed theoretically in~\cite{Bar_prl}) that  -- surprisingly -- it has a more marked monocentric aspect than London. In other words, the congestion in central places in New York is so large that introducing an efficient subway system is always better, even if it creates congestion at other points. Remarkably, these results on BC and on the existence of an optimal point are thus in agreement with a recent theoretical model of coupled transportation networks, where -- depending on the distribution of trip targets -- two regimes were observed: one in which the optimal coupling is trivially the maximum, an another one where a non-trivial optimal coupling exists~\cite{Bar_prl}.


\section{Discussion}


We have considered the effect of the coupling between two transportation layers on various quantities and we can summarise our results as follows. For quantities relating to quickest paths (interdependency, average quickest path duration), we observe a remarkable similarity between the two cities considered, suggesting the possibility of a universal behaviour requiring further study. This universality might originate in the fact that the quickest path can be seen as a sum of random variables, which inevitably leads to some sort of central limit theorem. This seems to be the case for the probability distribution of the quickest path time duration, that (once normalised) is a universal function for a reasonable range of subway speeds. More involved quantities such as the local outreach and the urban horizon also display a simple common behaviour that cannot be recovered using a back-of-the-envelope argument about the quickest path. This possible universality suggests that few parameters seem to govern the behavior of quickest paths, which could lead to many useful simplifications in more elaborated models that contain a large number of parameters. 
More data on more cities is however needed in order to validate this universal idea and to understand its origin.

We also studied the impact of the coupling on the spatial distribution of the betweenness centrality. We observe results in agreement with previous theoretical findings, with in particular the existence for London of an optimal subway velocity in terms of congestion. These results on spatial distribution of centralities can be also understood in the framework of another study showing how decentralising housing in London leaves room for commercial activities~\cite{Lev_2007}. Even if the direct causal link between land-use change and transportation network evolution is still not clear~\cite{King_2011}, our results seem to go in that direction. It would be interesting to confront our results with realistic models used by the transportation community, in particular when the subway speed is modified.  More generally, we believe that more empirical studies are however needed in order better to understand the complex coupling between land-use and the structure of multimodal networks. 

It thus seems to be clear that it is important to consider full multimodal, multi-layer network aspects in order to understand the behaviour of an urban transport system -- and thus understand the effects of transport on other features of interest. Even if these studies are still very theoretical, they show convincingly that reasoning with only one transportation mode can be extremely misleading, and that policy-makers cannot limit themselves to a single aspect of an urban system without risking making decisions that are locally correct but globally wrong.


\bigskip \bigskip 

\section{Data accessibility}

Original network data is available on Open Street Map~\cite{OSM}. The cleaned and topologically corrected networks are also freely available~\cite{figshare}.
They include ArchMap shape file (.shp) containing streets and underground networks with their adjacency lists. These files can be opened on any GIS platform.

\section{Competing financial interests}
The authors declare no competing financial interests.

\section{Author Contributions Statement}

SS wrote computational scripts and performed the computations on the multiplex networks. 
ES prepared the data and produced the maps. ES, SS, SD and MB designed research and wrote the paper.
 
\section{Acknowledgements}

ES thanks Bilal Farooq, Riccardo Scarinci, Michel Bierlaire, Sergio Porta and Luis Bettencourt for their suggestions at various stage of the research. ES and SS thank Sergio Porta for hosting us at his lab in Glasgow at the early phase of the project. MB thanks Riccardo Gallotti for discussions.

\section{Funding}

MB acknowledges funding from the European Commission FET-Proactive project PLEXMATH (Grant No. 317614). SS thanks the James S. McDonnell Foundation 21st Century Science Initiative - Complex Systems Scholar Award (grant 220020315) and the Scottish Informatics and Computer Science Alliance for financial support.

\section{Appendix: Total congestion}

\begin{figure}[ht]
\includegraphics[width=.4\textwidth]{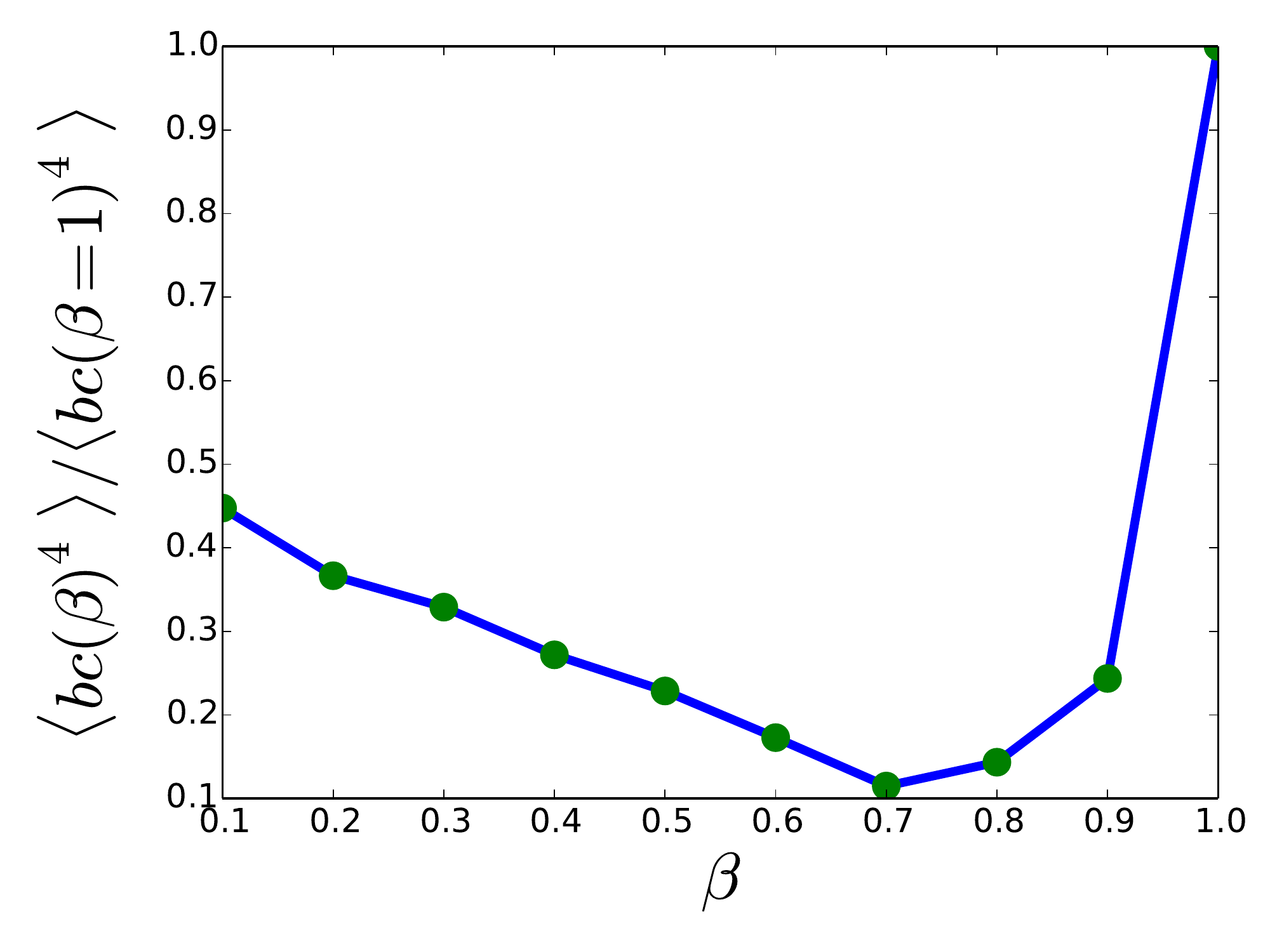}
\includegraphics[width=.4\textwidth]{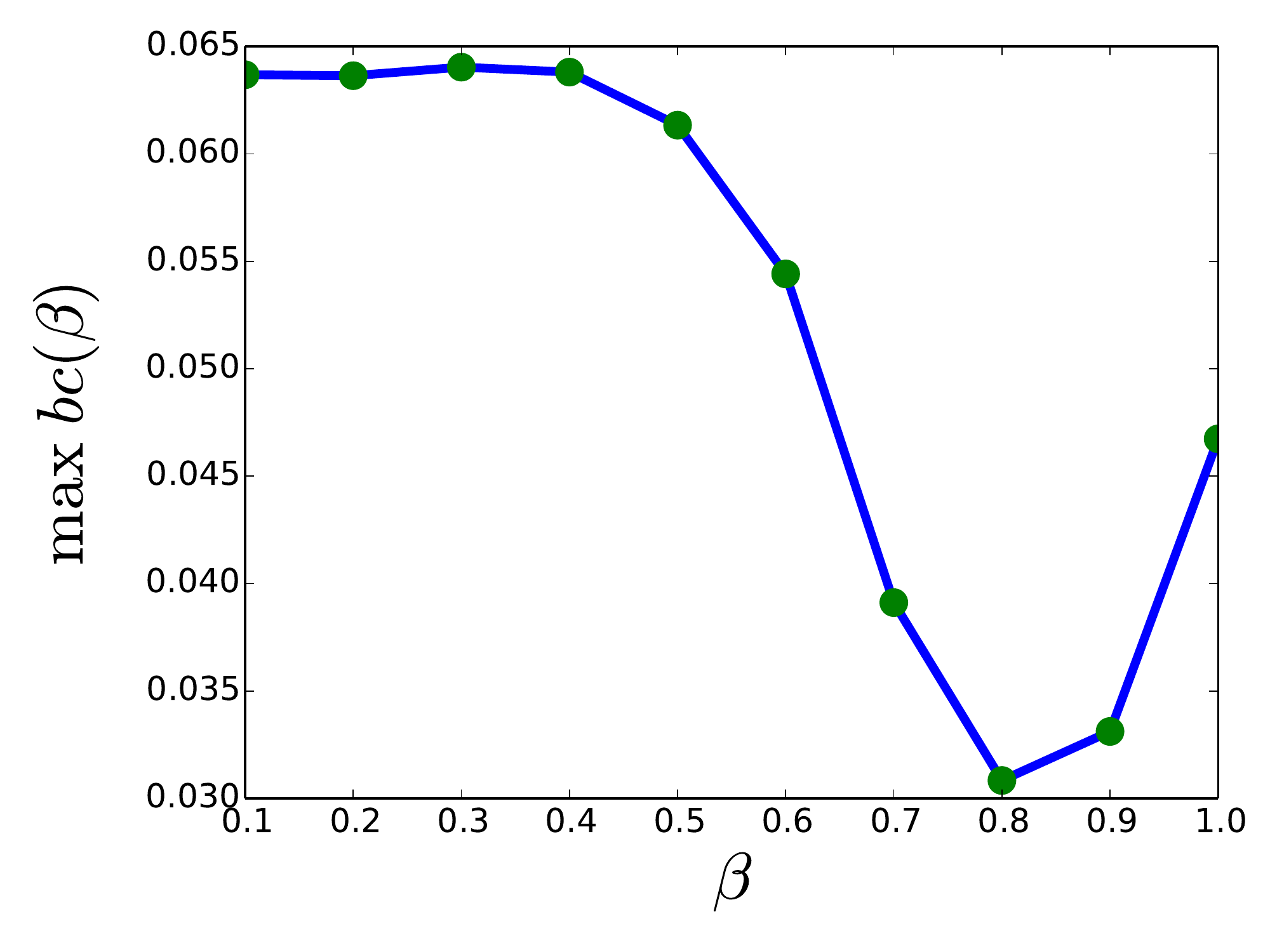}
\caption{Total (normalized) congestion computed for $\nu=4$ (left) and maximum congestion (right) as a function of $\beta$.} 
\label{fig:totalcon}
\end{figure}

We can choose as a measure of congestion the total (or equivalently the average) time spent on roads. We use the 
standard Bureau of Public Roads function \cite{Branston:1976}  that gives the time to go from $i$ to $j$ (separated by a distance $d(i,j)$ on a road with traffic $T(i,j)$ and capacity $c$
\begin{equation}
\tau_{ij}=\frac{d(i,j)}{v}\left[1+\left(
\frac{T(i,j)}{c}
\right)^\nu\right]
\end{equation}
where $v$ is the average velocity on this road and where the exponent $\nu$ is usually large $\nu\approx 4-5$.
In our model, the proxy for the traffic is given by the BC and the total congestion $C_T$ is thus approximately equal to (up to a constant factor independent from $\beta$)
\begin{equation}
C_T\propto\frac{\langle d\rangle}{v}\langle (bc)^\nu\rangle
\end{equation}
(we assumed that the length distribution of segments is peaked around its average $\langle d\rangle$).
In the figure \ref{fig:totalcon}, we plot this total congestion \-- normalized by its value at $\beta=1$ \-- as a function of $\beta$.
We observe on this figure a clear optimum which corresponds to the minimum of the Gini coefficient. Congestion is actually very sensitive to large values of the traffic and since the maximum has also a minimum (Fig.~\ref{fig:totalcon}) we can indeed expect that the larger the traffic heterogeneity and the larger the total congestion.

\bibliographystyle{prsty}

\end{document}